# New Type of Quantum Oscillations Stemmed From the Strong Weyl Fermions - 4*f* Electrons Exchange Interaction

Jin-Feng Wang[1,2], Qing-Xin Dong[1,3], Yi-Fei, Huang[1,3], Zhao-Sheng Wang[5], Zhao-Peng Guo[1,3], Zhi-Jun Wang[1,3,4], Zhi-An Ren[1,3,4], Gang Li[1,3,4], Pei-Jie Sun[1,3,4]*, Xi Dai[1,6,7]*, and Gen-Fu Chen[1,3,4]*

[1]*Institute of Physics and Beijing National Laboratory for Condensed Matter Physics, Chinese Academy of Sciences, Beijing 100190, China*

[2]*Henan Normal University, College of Physics, Xinxiang, Henan 453007, China*

[3]*School of Physical Sciences, University of Chinese Academy of Sciences, Beijing 100049, China*

[4]*Songshan Lake Materials Laboratory, Dongguan, Guangdong 523808, China*

[5]*Anhui Province Key Laboratory of Condensed Matter Physics at Extreme Conditions, High Magnetic Field Laboratory of the Chinese Academy of Sciences, Hefei, Anhui 230031, China*

[6]*Materials Department, University of California, Santa Barbara, CA 93106-5050, USA*

[7]*Department of Physics, The Hongkong University of Science and Technology, Clear Water Bay, Kowloon 999077, Hong Kong, China*

*E-mail: pjsun@iphy.ac.cn; daix@ucsb.edu; gfchen@iphy.ac.cn

**Abstract**

The interplay between magnetism and the topology of electronic band structure may generate new exotic quantum states. Here we report on a new type of quantum oscillations in the temperature dependent electrical resistivity and specific heat at a constant magnetic field in a polar magnetic Weyl semimetal (WSM) NdAlSi. These novel quantum phenomena arise from the destructive interference between quantum oscillations from the spin-split Fermi surfaces due to the strong Weyl fermions-4$f$ electrons exchange interaction combined with Rashba-Dresselhaus (RD) and Zeeman effects. Our findings pave a way to explore unprecedented quantum phenomena in 4$f$-electron based magnetic semimetals.

The ternary *R*-Al-*B* (*R*: Ce, Pr, Nd, Sm; *B*: Ge or Si) compounds, being candidates of Weyl semimetals, have drawn great interests due to their lacking both inversion and time-reversal symmetries, which offer a rare, unambiguous platform to explore exotic quantum phenomena arising from the interplay between magnetism, electronic correlations and topology [1-8]. Recent studies of NdAlSi revealed that the magnetic order in NdAlSi is driven by Weyl exchange interactions, which provides a concrete example of Weyl fermions promoting collective magnetism [9-10]. It is therefore not surprising that NdAlSi hosts a complex temperature-magnetic phase and a peculiar magneto-transport behavior [10-11]. Intriguingly, Weyl fermions survive in NdAlSi below or above $T_N$ under applied magnetic fields, suggesting a possible unconventional paramagnetic state that emerges from the suppression of magnetic order. Moreover, the energy splitting between the *f*-electron ground state doublet and four excited doublets is rather small, ∼ 4.1 meV, indicating that the excited state can mix easily into the ground state by applying magnetic fields [11]. Thus, the high field phase of NdAlSi might be the subject of particular interest.

In this work, we report the first experimental observation of new type of quantum oscillations in NdAlSi. We find that, both the electrical resistivity and specific heat oscillate clearly as a function of temperature at a constant high magnetic field, in addition to the conventional magnetic field dependent quantum oscillations at a constant temperature. Furthermore, we propose a phenomenological model to quantitatively account for the unique quantum oscillations observed herein, by including the spin-split Fermi surfaces due to the strong Weyl fermions-*f* electrons exchange interaction.

NdAlSi crystallizes in the noncentrosymmetric space group *I*4$_1$*md*. The body-centered tetragonal unit cell of NdAlSi is shown in Fig. 1(a) [11]. Figure 1(b) illustrates typical examples of the temperature-dependent resistivity at various magnetic fields (*T* < 100 K; *B* > $B_C$). All of them exhibit anomalous behaviors with oscillatory features mediated by magnetic fields. It can be seen from the figure that the peak position of magneto-resistivity anomaly shifts towards high temperature region accompanied by a singular enhancement of the amplitude with increasing magnetic field. The anomalous component, obtained by subtracting a smooth background from the experimental data, is plotted versus temperature in logarithm scale in Fig. 1(d). The most striking feature is that the spindle-like oscillations show a log-periodic-like temperature dependence. The spindles' features mirror the rise and decline of oscillations aptitudes and reflect the dramatic changes in the potential across a critical regime.

To get more insight into the origin of such phenomenon in NdAlSi, we performed detailed isothermal magnetoresistance (MR) measurements on NdAlSi, as well as a corresponding non-4$f$ reference compound LaAlSi. Figure 2(a) shows the SdH oscillations for NdAlSi at different temperatures, obtained by subtracting a smooth background from MR data in the $T > T_N$ region. Gradual variations of the oscillation spectra combined with phase inversions are clearly observed. It is quite surprising to observe such a very strong shift of the SdH oscillations' maxima (or minima) with temperature.

In order to further clarify the beating-like patterns of oscillations, we plot $\Delta\rho$ as a function of inverse field (1/$B$) in Fig. 2(b). As the results show, the phase inversions occur exactly at the positions of the nodes, i.e., the phase of the SdH oscillations remains the same between adjacent nodes and changes by $\pi$ after passing the nodal point. The beating effect observed in the quantum oscillations is usually due to the existence of two comparable frequencies with similar amplitudes due to level splitting, and the nodal positions are controlled by the difference of the two spin (up and down) sub-densities [12-15]. However, spin splitting zeros in our case are not perfect. Most strikingly, there is a large shift of the nodal position with temperature. As shown in Fig. S8 and Fig. S12, although the frequencies observed in NdAlSi and LaAlSi are similar, an interesting difference between them is visible. There is no obvious shift of the peak and valley positions of quantum oscillations for LaAlSi [16]. This fact suggests that while the 4$f$ electrons of Nd are essentially localized in the polarized paramagnetic (PPM) phase of NdAlSi, they would affect the quantum oscillations substantially through exchange coupling with the conduction electrons, i.e. Weyl fermions.

Figure 2c displays the oscillation component as a function of 1/*B* at selected temperatures. These competing oscillations are 180° out of phase over all measured field range, where the temperature intervals are comparable to the periods observed in the temperature dependent resistivity. Alternative to the directly measured magneto-resistivity, we can also plot the resistivity versus temperature curves at selected magnetic fields using the data taken from the above isothermal MR measurements. As shown in Fig. 2(d), the red line is the extracted magneto-resistivity, and the open circles are the directly measured one. There is no doubt that the two curves are identical, demonstrating clearly that the unexpected temperature dependent resistivity oscillations arise from the destructive interference between two quantum oscillations from the Fermi surface splitting, accompanied by a strong temperature and magnetic field dependent phase inversion.

Let's consider here the oscillatory magnetoresistance using the 3D Lifshitz-Kosevich (LK) formula with arbitrary band dispersion [17-18].

$$\Delta\rho = \left(\frac{B}{2F}\right)^{1/2} R_T R_D R_S \cos\left[2\pi(\frac{F}{B} + \gamma - \delta)\right], \quad (1)$$

where *F* is the oscillation frequency, and $\gamma = \frac{1}{2} - (\varphi_B/2\pi)$ is the Onsager phase factor taking the value *γ* = 0, 1 (or *γ* =1/2) for the nontrivial (trivial) Berry phase $\varphi_B$. The offset δ is a phase shift determined by the dimensionality, *δ* = 0 (or ±1/8) for the 2D (or 3D) system, $R_T = \alpha T\mu/[B\sinh(\alpha T\mu/B)]$ and $R_D = \exp(-\alpha T_D\mu/B)$ are thermal and Dingle damping factors due to Landau level broadening, caused by the finite-temperature effect on the Fermi-Dirac distribution and the electron scattering, respectively. $R_S = \cos\left(\pi\frac{\Delta E}{\hbar\omega_c}\right)$ is spin damping factor due to Zeeman splitting, where $\hbar\omega_c$ is the Landau level separation energy and *ΔE* is the energy splitting of each Landau level. In the conventional linear Zeeman effect, *ΔE* = *gμ$_B$B* (*g*: *g*-factor), and thus $R_S = \cos(\pi gm^*/2m_e)$ becomes independent of *B*. So called "spin zeros" will appear as $gm^*/m_e = 2n+1$ (n is integer, *m*$^*$ is the effective mass of electrons), and the possible value of *g* factor can be then obtained [17]. Note that the above spin factor $R_S = \cos(\pi gm^*/2m_e)$ is obtained for free nonrelativistic electrons with parabolic band dispersion.

While for a Dirac cone, the energies of Landau levels, $E_n = \sqrt{2\hbar e v_F^2 nB}$, are not equally spaced ($v_F$ is the Fermi velocity of the system with linear dispersion), and the energy separations are dependent on $\sqrt{B}$ [19-20]. Hence the $R_S = \cos\left(\pi \frac{\Delta E}{\hbar \omega_c}\right)$ will be field dependent, and leads to a beating pattern in oscillations at half-integer values of $\Delta E/\hbar\omega_c$.

In NdAlSi, in addition to Zeeman splitting, the asymmetric RD interaction, driven by spin-orbit coupling (SOC) in the absence of inversion symmetry, also induces a density imbalance between spin-up and spin-down electronic sub-bands even at $B = 0$. However, both Zeeman and RD effects cannot explain why the observed nodal positions shift with temperature. Therefore, an additional exchange splitting [21-23], stemming from exchange interaction between the conduction electrons (i.e. Weyl fermions) and the localized 4$f$ electrons of Nd [9-10], appears to be crucial and results in the changing of band splitting with temperature. Usually, the exchange interaction, acts as an effective field that results in populations of spin-up and spin-down quasiparticles that are no longer equal, which prevents a regular interference between oscillations from the two spin bands and might lead to the absence of a perfect spin splitting zero in the oscillation patterns. Such feature is consistent well with the observation in NdAlSi. So the total spin splitting energy $\Delta E$ in NdAlSi under external magnetic field should to be a combination of RD spin splitting, Zeeman splitting and $c$-$f$ exchange splitting energies. To the best of our knowledge, this is the first observation of the occurrence of multiple effects in topological quantum materials.

In such case, the spin split zeros will appear as $g_{eff} m^*/m_e = 2n+1$, if we introduce an effective $g$-factor, $g_{eff} = \frac{1}{\mu_0 \mu_B}\left(\frac{dE_{F\uparrow}}{dH} - \frac{dE_{F\downarrow}}{dH}\right)$. Unfortunately, due to the complicated structure of spin nodes, as well as the limited number of spin nodes available, it is not yet possible to describe the variation of the spin damping factor in NdAlSi, although the $R_S$ term provides also a handle for estimating the exchange interaction. An effective assessment of the $R_S$, and the resultant $g$-factor (large and tunable) might be a big challenge, which remains a subject of further theoretical and experimental studies.

Alternatively, we herein develop a simple phenomenological model to capture the temperature dependent oscillations without going into the microscopic details of the damping factor in NdAlSi, considering that the exchange interaction is determined by the magnetization. In general cases, the LK formula (1) could be written as,

$$\Delta\rho = A(B,T)\cos\left[2\pi\left(\frac{F}{B}+\varphi\right)\right] \quad (2),$$

where $A(B, T)$ is the SdH oscillation amplitude (governed by both thermal broadening and impurity scattering). Due to the Fermi surface splitting, the SdH frequency $F$ is split into two comparable frequencies, $F_+$ and $F_-$, and thus the conventional LK equation (2) should be modified as the following:

$$\Delta\rho = A_1(B,T)\cos\left[2\pi\left(\frac{F_+}{B}\right)+\phi_+ + \pi\mathcal{G}_+ + \xi\right] + A_2(B,T)\cos\left[2\pi\left(\frac{F_-}{B}\right)+\phi_- + \pi\mathcal{G}_- + \xi\right] \quad (3),$$

where $\mathcal{G}_+$ and $\mathcal{G}_-$ present the spin states arising from the Zeeman splitting, $\xi$ is the field independent phase stemming from the RD splitting. For simplicity, considering $\mathcal{G}_+ + \mathcal{G}_- = 0$, and $\mathcal{G} = \mathcal{G}_+ - \mathcal{G}_- = gm^*/2m_e$ and the amplitude and frequency are spin independent, summing over the equation (3) yields:

$$\Delta\rho = 2A(B,T)\cos\left[2\pi\left(\frac{F}{B}+\phi'\right)\right]\cos\left(\pi\left(\frac{\Delta F}{B}+\mathcal{G}\right)\right) \quad (4),$$

where $F = (F_+ + F_-)/2$, $\Delta F = F_+ - F_-$. Note that the frequency difference $\Delta F$ between the two opposite spin electron oscillations depends on the exchange splitting energy $E_{ex}$, which is generally proportional to the magnetization $M$ in magnetic metals [24-25]. So we obtained,

$$\Delta\rho = 2A(B,T)\cos\left[2\pi\left(\frac{F}{B}+\phi'\right)\right]\cos\left(\pi\left(\frac{\alpha M(B,T)+\Delta_0}{B}+\mathcal{G}\right)\right) \quad (5),$$

in which $\alpha$ is the exchange interaction constant, and $\Delta_0$ is the exchange splitting at $T = 0$, and $M(B, T)$ is explicitly a function of both temperature and magnetic field. This approach gives the possibility of including the exchange interaction of electrons with magnetic ions into a simple band structure model. The last term, $\pi(\frac{\alpha M(B,T)+\Delta_0}{B} + \mathcal{G})$, can lead to the beating patterns in oscillations at half-integer values, which is consistent with the observation that the nodal positions shift with temperature variation in NdAlSi.

Therefore, within the model above, we can use the magnetization data *M/B*, which traces the spin-polarized *f* electrons under the influence of the conduction electrons-*f* electrons (*c-f*) exchange interactions to reproduce the temperature dependent resistivity curves at various magnetic fields with the resulting expression (5). Due to the constant value of *B*, the expression (5) could be simplified as: $\Delta\rho = \cos(\frac{\alpha'M}{B} + \frac{\Delta'_0}{B} + \delta)(-AdM/dT)$, where *A*, $\alpha'$, $\Delta'_0$, $\delta$ are constant parameters. The last term, $-AdM/dT$, is the correction of the amplitude, due to the changes we observed in *A(B,T)* are proportional to the changes in $-M(B,T)$. As shown in Fig. 1(d), the oscillatory resistivity data are well reproduced by the expression, which confirms further that the observed novel oscillations arise from the dynamical coupling of the Weyl fermions and *f* electrons in NdAlSi. From the obtained $\alpha'$ (0.025±0.001), $\Delta'_0$ (24±2), and the corresponding effective cyclotron mass *m*\* (0.098 $m_0$, $m_0$ is the free-electron mass, see Fig. S8), we may evaluate the strength of the 4*f*-Weyl fermions exchange interaction in NdAlSi, which is found to be about 0.16 eV. It is very close to the energy difference (~0.15 eV) between spin-up and spin-down bands at the Fermi level in PPM state by DFT calculation (Fig. 2(f)). That value is also comparable to the strength of the 4*f*-*s* interaction (0.2 eV) calculated theoretically by Harmon and Freeman [26], but is smaller than the value (0.5 eV) of that between 4*f* and *d* elements. Taking the same obtained parameters, we have also simulated the conventional oscillating magneto-resistivity $\Delta\rho$ using the expression (5), and compared with the experimental curves as shown in the inset of Fig. 2(d). The good agreement between the theoretical and experimental curves proves that origin of beating patterns of the SdH oscillations is due to the interplay of RD, Zeeman and *c-f* exchange splitting.

Fascinatingly, an oscillatory feature was also observed in the magnetic field dependent specific heat for NdAlSi, as shown in Fig. 3(a). Such an effect, caused by quantum oscillations of the density of states, being not directly related to electronic scattering, can give important and reliable information on particularities of the band structure and the density of electron states. Here the most remarkable feature is that, unlike the SdH oscillations, the specific heat $C$ shows only few oscillations in the intermediate field region before the system becomes fully polarized. After subtracting a smooth background component, $C_{osc}$ oscillations with a temperature independent frequency of 45 T is obtained, as shown in inset of Fig. S9(d). One can see that the specific heat oscillation minima correlate precisely with the maxima peaks of SdH oscillations in MR (see inset of Fig. 3(b)), due to the oscillatory behavior in specific heat characterized by the sine term [17].

We have also performed measurements of the temperature dependence of specific heat at various magnetic fields, as shown in Fig. 3(c). The most intriguing feature obtained here is the saddle-like structure, a superimposed oscillation due to the quantum oscillations of the density of electron states. It should be pointed out that, the oscillatory component of the total heat capacity is very small due to the huge background arising from CEF splitting. To see the temperature variation of $C_{osc}$ more clearly, we plot $C_{osc}$ as a function of temperature on logarithm scale at selected magnetic fields in the Fig. 3(d), after subtracting the contribution of the lattice and crystal field effect. An oscillatory character of the electronic specific heat with respect to magnetic field is clearly demonstrated.

Remarkably, the low temperature heat capacity data can be fitted well with an expression for activated behavior ($Ae^{\frac{-\Delta_\varepsilon}{T}}$) in the high field phase (see Fig. 4(a)), in which the spins are entirely polarized by the external fields with ferromagnetic like correlations, leading to the opening of a magnetic excitation gap [27]. It reminds us that the drastic decrease of low temperature resistivity under high fields may have the same origin. We also estimate the excitation gap $\Delta_\varepsilon$ by $\rho = Ae^{\frac{-\Delta_\varepsilon}{T}} + \rho_0$, and plotted it versus the corresponding magnetic field $B$ in Fig. 4(b), along with gap size deduced from the heat capacity measurements. The gap sizes $\Delta_\varepsilon$ for them are found to differ significantly, but both tend to go to zero at finite magnetic field ($B_c^* \sim 3$ T) rather than at zero field, as would be excepted for a field-polarized paramagnetic ground state (see Fig. 4(b)). It should also be remarked that the critical field $B_c^*$ value is slightly different from the critical field $B_c \sim 5.5$ T , indicating that the polarization occurs before the suppression of AF order, which is consistent with the scenario that the polarization destroys the antiferromagnetism in NdAlSi [11].

Temperature-field phase diagram for NdAlSi is displayed in Fig. 4(c) for $B//c$, based on the $M/B$ vs. $T$ and $C$ vs. $B$ data. It shows clearly that the paramagnetic phase at low magnetic fields and high temperatures becomes polarized with increasing field strength and decreasing temperature, therein a cross-over from a paramagnetic to a ferromagnetic-like polarized state emerges. Amazingly, the temperature dependent oscillations occur only in the crossover regime, as discussed above. In particular, the maximum amplitude of the spindle-like oscillations appears at a critical temperature, $T^*$, which are just located at so called "boundary" between the paramagnetic and the polarized paramagnetic (PPM) phases. This demonstrates again that all the unconventional behaviors observed in NdAlSi are strictly linked to the existence of $c$-$f$ exchange splitting mechanism. During the magnetic polarization process of the localized 4$f$ electrons, the two opposite spin states are further mixed by RD interactions, Zeeman and CEF effects at the Fermi levels.

In summary, we have observed a new type of quantum oscillations in the temperature dependence of resistivity and specific heat at a constant high magnetic field in magnetic Weyl semimetal NdAlSi. Such oscillations appeared in the field induced crossover regime from PM to PPM is particularly intriguing. We have proposed a phenomenological model to address these exotic quantum oscillations well by considering the exchange splitting effect that the conduction band in a polarized state is split into two opposite spin sub-bands through the strong exchange interaction with the localized $f$ electrons, additional to the Zeeman and RD spin splittings. Our study opened up a new research avenue in the subject of $f$-electron based topological systems. Continuing exploration of such materials may offer an increased possibility for realizing new and intriguing quantum phenomena.

This work is supported by the National Natural Science Foundation of China (Grant No. 11874417, U1504107), the Strategic Priority Research Program (B) of Chinese Academy of Sciences (Grant No. XDB33010100). We would like to thank Li-Yu, Wen-Liang Zhu, Yi-Yan Wang, Meng Lv and Ya-Dong Gu for helpful discussions. J. F. Wang, Q. X. Dong contributed equally to this work.

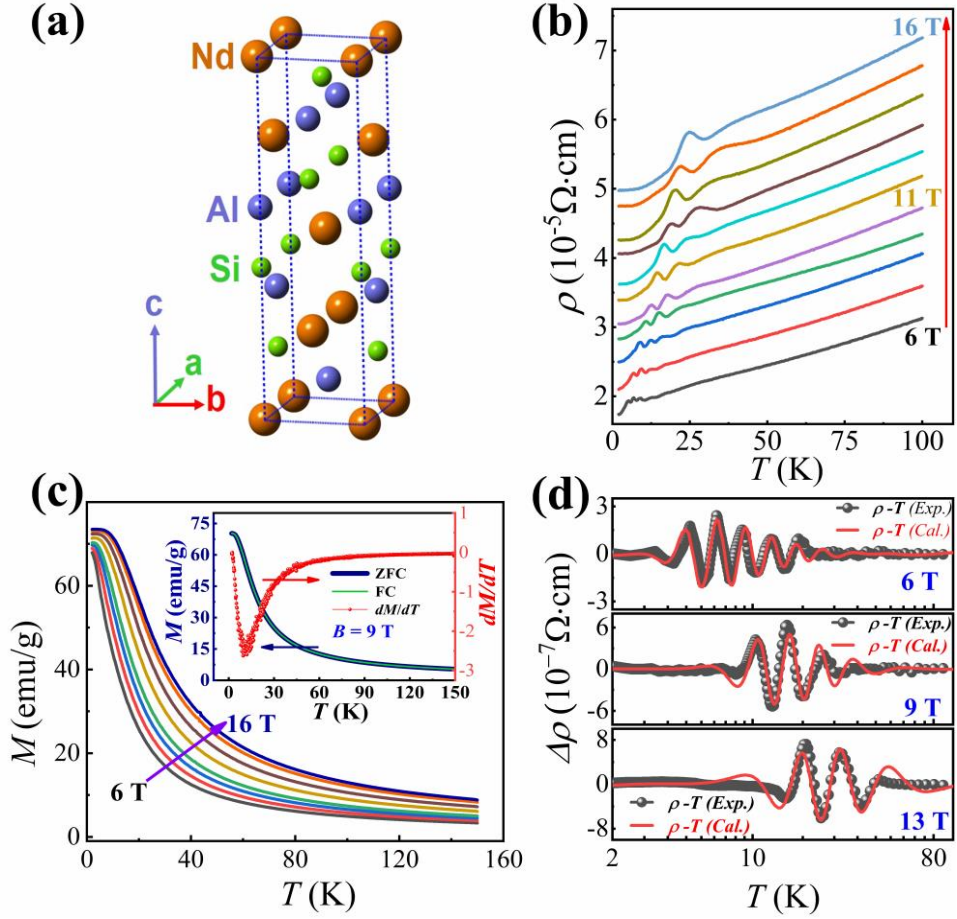

FIG. 1. Non-centrosymmetric crystal structure, high field magnetization and temperature dependent magneto-resistivity oscillations in NdAlSi. (a) Crystal structure with a space group of $I4_1md$ (tetragonal LaPtSi-type). (b) Temperature dependent resistivity $\rho(T)$ under external magnetic fields. $\rho(T)$ curves are vertically shifted ($0.35 \times 10^{-5} \Omega \cdot cm$) for clarity. (c) Temperature dependent zero-field-cooled (ZFC) magnetization at various magnetic fields. The magnetization $M$ tends to saturate with lowering temperature, indicating the polarization of $Nd^{3+}$ moments. Inset: The $M_{ZFC}$ (blue) and $M_{FC}$ (green) curves overlap each other in the whole temperature regime for $B = 9$ T, excluding the possibility of a ferromagnetic phase transition. The red line represents the derivative $dM/dT$ for $M_{ZFC}$. (d) Extracted $\Delta\rho$-$T$ oscillations (black dotted lines) for $B = 6$ T, 9 T and 13 T, demonstrating a log-periodic-like resistivity oscillations. The red lines are the fitting results based on a simplified equation (5) in the main text: $\Delta\rho = \cos\left(\frac{\alpha'M}{B} + \frac{\Delta'_0}{B} + \delta\right)(-AdM/dT)$. Note also that, all the experimental resistivity, magnetoresistance and specific heat data in the present work were measured for $B$//c.

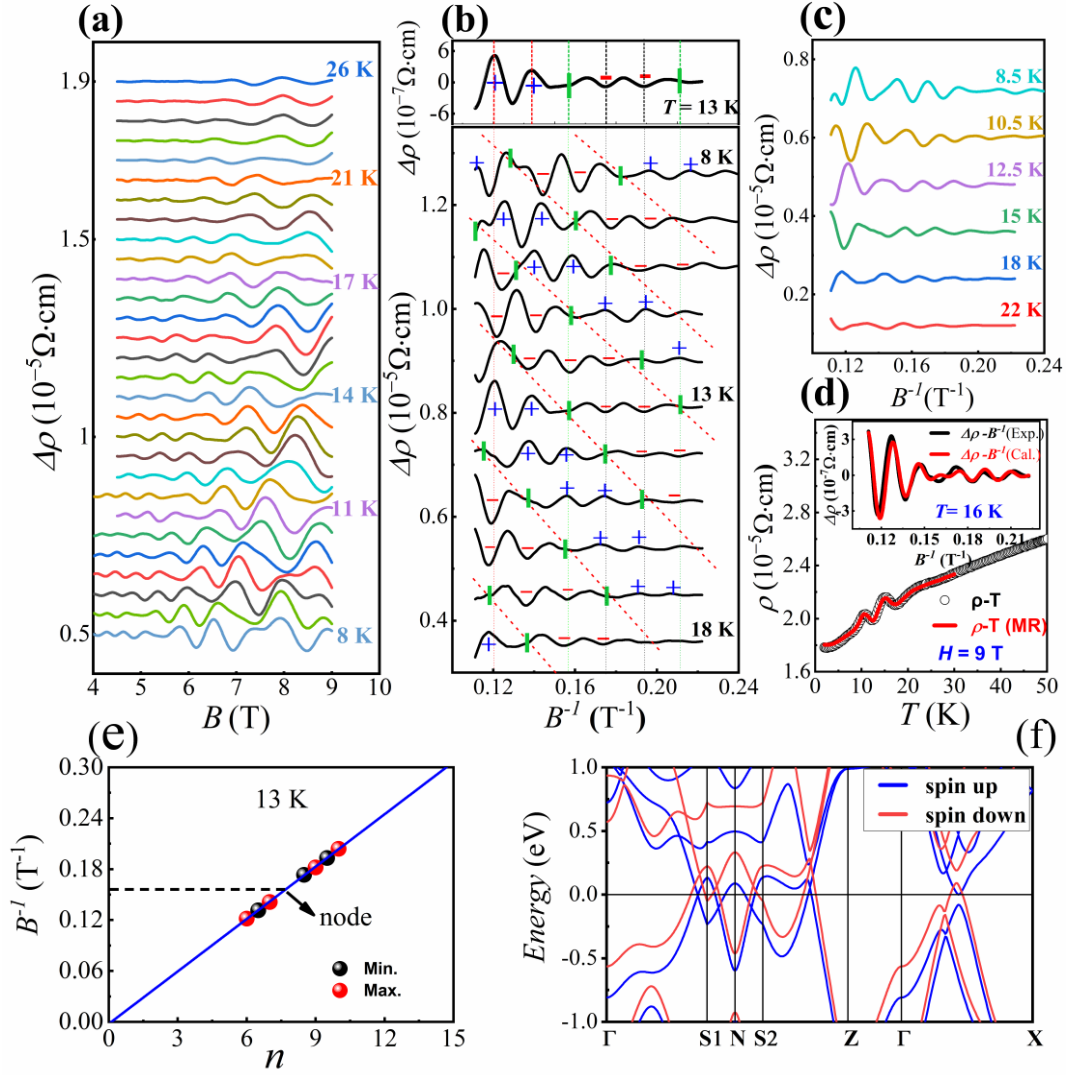

FIG. 2 Exploration of the $\Delta\rho$-$T$ oscillations' origin based on SdH oscillations analyses. (a) Extracted SdH oscillations measured at a paramagnetic wide temperature range. (b) SdH oscillations plotted as a function of $1/B$ for 8~18 K with a 1 K increment. The phase inversion positions (so called 'nodes') were labeled as green vertical bars. (c) Oscillatory $\Delta\rho$ versus $1/B$ at the selected temperatures, displays a 180° phase inversion with respect to each other. (d) The directly observed $\rho(T)$ curve matches well with the extracted one from the MR data f. Inset: The observed SdH oscillations and its simulation based on the expression (4) in the text using the $M$-$H$ data. The corresponding SdH curves in (a), (b), (c) are vertically shifted of 0.05, 0.09, 0.12 for clarity. (e) Landau level index plot $1/B$ versus n. The linear extrapolation of the plot yields the value of Onsager phase to be 0.11, evidencing the presence of nontrivial π Berry phase arising from 3D Weyl electrons in the PPM phase. (f) The band structures of NdAlSi in the PPM state without spin-orbit coupling.

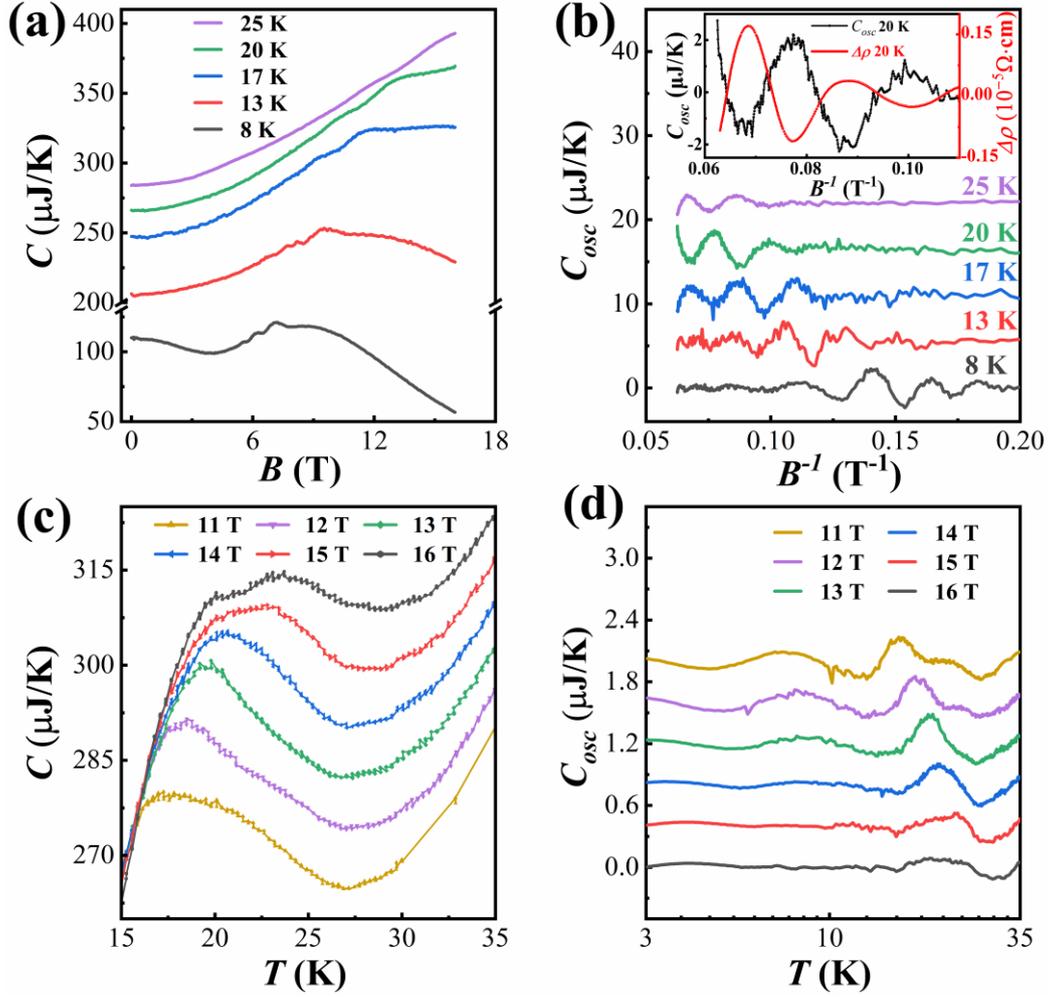

FIG. 3. The specific heat oscillations in NdAlSi. (a) Magnetic field dependence of the heat capacity (*C-B*) measured at selected temperatures. (b) Extracted oscillating component $C_{osc}$ of heat capacity plotted as a function of $1/B$. The $C_{osc}$ oscillations exhibit a 180° phase change compared with that of $\Delta\rho$, as shown in the inset. (c) *C-T* curves under several magnetic fields. The heat capacity shows a broad hump (CEF anomaly) that increases in temperature with increasing magnetic field. (d) Extracted $C_{osc}$ oscillations *vs.* temperature for the selected magnetic fields. The $C_{osc}$ curves in b and d are vertically offset of 5.5 μJ/K and 0.4 μJ/K for clarity.

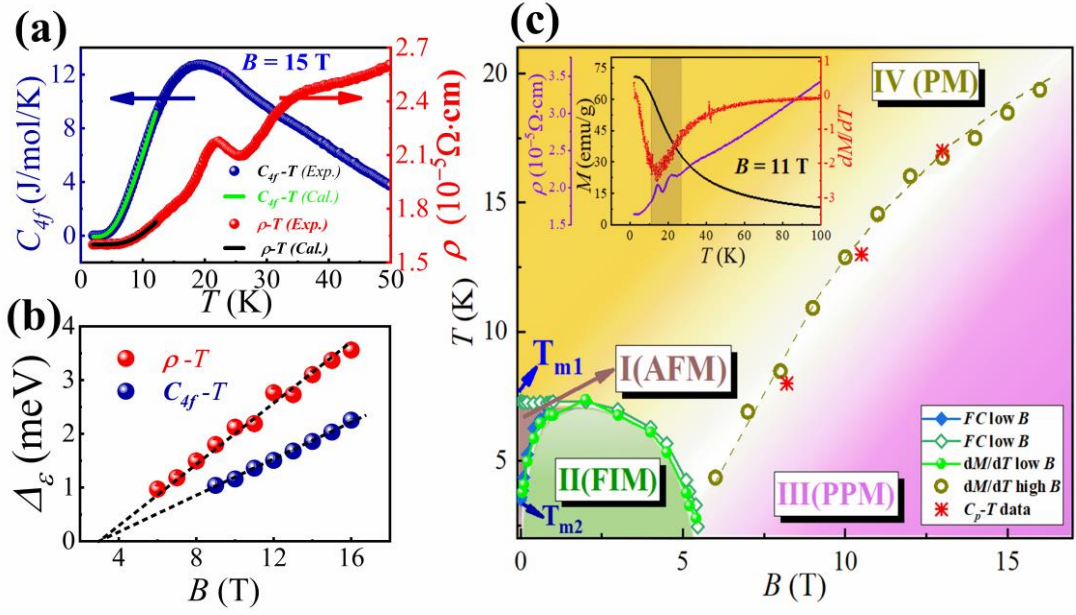

FIG. 4. Low temperature magnetic excitations and magnetic phase diagram with applied field //c-axis. (a) Typical $C_{4f}$-$T$ and $\rho$-$T$ curves measured at $B$ = 15 T ($C_{4f}$ = $C_{NAS}$-$C_{LAS}$). The green and red lines are the fitting results with the expressions $C_{4f} = Ae^{-\frac{\Delta\varepsilon}{T}}$ and $\rho = Ae^{-\frac{\Delta\varepsilon}{T}} + \rho_0$, respectively. (b) The calculated magnetic excitation gap $\Delta_\varepsilon$ (meV) plotted as a magnetic field dependence, which tend to merge at $B_c^* \sim 3$ T. (c) Magnetic field–temperature phase diagram of NdAlSi based on the corresponding magnetization and specific heat data. The high-field part of the phase diagram is characterized by a crossover at a critical temperature $T_p$, defined here as an inflection point of the $M(T)$ vs. $T$ curve measured at various magnetic fields, between the paramagnetic phase and the high-field polarized paramagnetic phase. Inset: Low temperature $\rho$, $M$ and $dM/dT$ for $B$ = 11 T. The novel oscillations appear only in a limited temperature range at which the magnetization undergoes a drastic change, see the shadowed region in the inset.